\documentclass[twocolumn,showpacs,preprintnumbers,amsmath,amssymb,pra]{revtex4}

\usepackage{graphicx}
\usepackage{dcolumn}
\usepackage{bm}

\newcommand{\vv}{\overleftrightarrow}
\newcommand{\de}{\delta\epsilon}
\newcommand{\te}{\tilde\epsilon}
\newcommand{\conv}{\otimes}
\newcommand{\ii}{\imath}

\begin{document}


\title{On the universality of the Discrete Nonlinear Schr\"odinger Equation}

\author{Andrea Fratalocchi}
 \email{fratax@gmail.com}
\affiliation{
Museo Storico per la Fisica e Centro studi e ricerche ''Enrico Fermi'', Via Panisperna 89/A, I-00184, Roma, Italy\\
Research center CRS-SOFT, c/o University of Rome ''La Sapienza'', I-00185, Roma, Italy
}

\author{Gaetano Assanto}
 \email{assanto@uniroma3.it}
\affiliation{
Nonlinear Optics and OptoElectronics Lab (NooEL),\\
INFN and CNISM, University ROMA TRE,
Via della Vasca Navale 84, 00146, Rome, Italy
}

\date{\today}

\begin{abstract}
We address the universal applicability of the discrete nonlinear Schr\"odinger equation. By employing an original but general top-down/bottom-up procedure based on symmetry analysis to the case of optical lattices, we derive the most widely applicable and the simplest possible model, revealing that the discrete nonlinear Schr\"odinger equation is ``universally'' fit to describe light propagation even in discrete tensorial nonlinear systems and in the presence of nonparaxial and vectorial effects.
\end{abstract}


\maketitle

\section{Introduction}
\label{intro}
The formulation of problems in Physics  results essentially in theoretical models able to describe the properties of a class of phenomena. Typically this consists of deriving a set of integro-differential equations, stemming from fundamental laws and shaving off all the assumptions which do not affect the predicted observables. The latter approach, known as Ockham's razor (OR) and summarized by the Latin sentence "entia non sunt multiplicanda praeter necessitatem" (i.e. "entities should not be multiplicated beyond necessity"), is commonly employed to discriminate between equally explanatory theories or competing systems of hypotheses. Furthermore the major goal of theories in physics should be the derivation of ``universal'' models, based on the conjecture that various phenomena are ruled by common characteristic principles, thereby allowing one to apply the same description to events taking place in  different areas/fields. The  question  arising naturally is then: how can we be sure that a model really conforms to the Ockham's razor or, equivalently, how can we guarantee that it entails both universal applicability and importance (see e.g. \cite{bf_um_chena,bf_um_chamsa,bf_um_rollia,bf_um_taniua,WII})? "Universality" should be intended here as the \emph{model} counterpart of ``universality'' in statistical mechanics, the latter based on the concept that properties exist for and apply to a large class of systems independently of their dynamical details (see e.g. \cite[Chapter 21]{TOCS} and \cite{sm_rg_kadana,sm_um_lawlaa,sm_um_celanaa,sm_um_astroa,sm_un_weissa,sm_um_bendea,sm_um_araia,sm_um_paczua,RG}).\\ 
Following the pioneering numerical experiments of Fermi, Pasta and Ulam on anharmonic lattices \cite{sol_bf_fermia,CPOEF}, the study of  energy dynamics in nonlinear discrete systems has attracted growing attention in recent years. Thereby, two major theories of the last half-century, namely the theory of integrable systems and the theory of chaos, have been specialized to account for discreteness \cite{SATIST,DACNSS,OTT,DC}. In this framework the nonintegrable discrete nonlinear Schr\"odinger equation (DNLS) has received significant interest \cite[and references therein]{dsol_by_scotta,dnls_eilbea}. The DNLS is frequently encountered in several areas of science, including biology \cite{wa_dav_davyda}, solid state physics \cite{df_ds_kenkra}, Bose-Einstein condensates \cite{wa_BEC_Tromb} and nonlinear optics \cite{dis_ChrisAA}, the latter being particularly relevant in view of the mature technology available for optical waveguides.  \cite{rev_wa_FleisA,rev_wa_CristA,OSFFTPC,SS,wa_NLC_FratTA,rev_wa_TrillA}. Despite its relevance the ``universality'' of this model has never been addressed: in nonlinear optics for instance the DNLS is derived assuming a paraxial regime, scalar Kerr nonlinearities and first order perturbations (see e.g. \cite[Chapter 13]{YAR}, \cite[Chapter 11]{OSFFTPC}, \cite[pages 269-290]{SS}, and \cite{dis_ChrisAA, rev_wa_TrillA}). Such hypotheses left a number of open issues concerning large nonlinear responses, large index perturbations (e.g., in deep gratings and in photonic crystals \cite{OPOPC}), vectorial and nonparaxial effects. The last, a subject of considerable interest in continuous media \cite{bf_npx_ciatta,bf_npx_manasa,bf_npx_gadia,bf_dsnpx_ciattaa,bf_npx_esareaa}, is of particular relevance in nonlinear optics owing to the recent progress in nanotechnology and the realizability of optical structures with subwavelength features. Before proceeding, however, we want to state clearly an operative definition of universality for physical models. \emph{ By "universality" we mean the widest applicability of a model in describing the largest family of phenomena and the various aspects of the dynamics they encompass}. We aim hereby at verifying the universality of the DNLS equation in nonlinear optics, adopting an approach of general applicability. We employ the language of symmetries and Lie transformation groups. They provide a powerful approach to the analysis , as witnessed by the large number of examples ranging from pure mathematics to chemistry, biology, optics, thermodynamics, solid state physics, quantum mechanics and robotics \cite{AOLGTDE,SADE,SAIMFDE,GTAIAIP,GTAIATPP,rb_sa_liua}. In particular symmetry methods and Lie transformation groups ---originally developed by Sophus Lie in the 19th century \cite{SLDIP}--- provide a systematic avenue to study nonlinear differential equations \cite{SAIMFDE,AOLGTDE}.  
Our analysis develops along a two-fold \emph{top-down/bottom-up} approach: in the first stage we perform a symmetry reduction of the problem stemming from fundamental laws and in the most widely applicable case, i. e. full vectorial Maxwell's equations in the nonparaxial regime and in the presence of a tensorial nonlinearity. In the second phase we proceed via Lie symmetrie to solve a classification problem and to assemble the simplest equation sharing the properties of the model derived in the previous step. In comparison with existing literature \cite{dis_ChrisAA,OSFFTPC,SS,rev_wa_TrillA} we remark that our approach deals with the large class of cases including nonparaxial vectorial propagation as well as a tensorial nonlinear response. This two-fold procedure yields the most widely applicable model with the simplest (i.e. the most elementary) structure possible: the latter is ensured by the classification (bottom-up analysis) and the former by the reduction step (top-down analysis). Furthermore it satisfies two fundamental requirements of predictive sciences: Ockham's razor and the principle of noncontradiction, the latter stated mathematically. We reveal that the DNLS is far more important than known to date because it is the simplest ``universal'' model of discrete nonlinear energy propagation, even in the presence of nonparaxial and vectorial effects, e.g. in optical lattices.
\section{Symmetry reduction analysis}
We begin by modeling the propagation of electromagnetic waves in a nonlinear medium using the action integral $\mathcal{I}=\int \mathcal{L}d\mathbf{r}dt$ and the Lagrangian:
\begin{equation}
\label{sylag0}
  \mathcal{L}=(\mathbf{d}_0+\frac{\mathbf{p}^{\mathrm{NL}}}{2}\conv)\mathbf{e}-\mathbf{h}\cdot\mathbf{b}
\end{equation}
with fields $(\mathbf{e},\mathbf{d}_0,\mathbf{h},\mathbf{b})$ defined by  potentials $(\phi,\mathbf{a})$:
\begin{align}
\label{gff0}
&\mathbf{e}=-\nabla\phi-\frac{\partial}{\partial t}\mathbf{a}=\vv\epsilon^{-1}\mathbf{d}_0, &\mathbf{b}=\nabla\times\mathbf{a}=\vv\mu\mathbf{h}
\end{align}
the nonlinear polarization $\mathbf{p}^{\mathrm{NL}}$ describing a generic nonistantaneous tensorial Kerr perturbation of components $p^{\mathrm{NL}}_i=\chi^{(3)}_{ijkl}\conv e^j\conv e^k\conv e^l$, $i,j,k,l\in[1,2,3]=[x,y,z]$ (assuming Einstein summation over repeated indices) with $\chi^{(3)}(\mathbf{r};t)_{ijkl}$ the nonlinear susceptibility and $\conv$ a convolution operator such that $f\conv a\conv b\conv c=\iiint d\xi_1 d\xi_2d\xi_3 f(\mathbf{r};t-\xi_1,t-\xi_2,t-\xi_3)a(\mathbf{r},\xi_1)b(\mathbf{r},\xi_2)c(\mathbf{r},\xi_3)$ \cite{NO}. The Euler-Lagrange (EL) equations are derived as conservation laws through Noether's theorem \cite{nt_bf_howara} by exploiting the variational symmetries generated by the basis $\{\mathbf{v}_1=\partial/\partial \phi,\mathbf{v}_2=\sum_j\partial/\partial a_j\}$ ($j\in[x,y,z]$) \cite[see e.g.][Chapter 4, page 276]{AOLGTDE}. We obtain:
\begin{align}
  \label{sselym0}
  &\partial_j\bigg(\frac{\partial \mathcal{L}}{\partial u^j_\alpha}\bigg)=0, &(\alpha\in [1,4])
\end{align}
with the four-element potential $\mathbf{u}=(\phi,a_x,a_y,a_z)$, $\partial_j=\partial/\partial j$ and $u^j_\alpha=\partial_ju_\alpha$. The EL equations (\ref{sselym0}), together with (\ref{gff0}), yield Maxwell's equations:
\begin{align}
 &\nabla\times\mathbf{h}=\frac{\partial}{\partial t}\mathbf{d}, \\
\label{edcm0}
&\nabla\times\mathbf{e}=-\frac{\partial}{\partial t}\mathbf{b},
\\
&\nabla\cdot\mathbf{b}=0,\\
&\nabla\cdot\mathbf{d}=0,
\end{align}
with $\mathbf{d}=\mathbf{d}_0+\mathbf{p}^{\mathrm{NL}}$. In a one-dimensional lattice \cite{dis_ChrisAA} the dielectric tensor is periodically modulated (with period $\Lambda$) along one direction, say $y$, and can be expanded as:
\begin{align}
\frac{\vv\epsilon}{\epsilon_0}\equiv\vv\epsilon'=\sum_n\vv\te(x,y-n\Lambda)=\vv\te(\mathbf{r})+\vv{\Delta\epsilon}(\mathbf{r})  
\end{align}
with $\vv\te(\mathbf{r})$ defining the \emph{canonical} structure and $\vv{\Delta\epsilon}(\mathbf{r})$ an arbitrary linear perturbation. We then Fourier transform the potential:
\begin{align}
  \begin{bmatrix}
\phi(\mathbf{r};t)\\
\mathbf{a}(\mathbf{r};t)
\end{bmatrix}=\int d\omega\begin{bmatrix}
\Phi(\mathbf{r};\omega)\\
\mathbf{A}(\mathbf{r};\omega)
\end{bmatrix}
\exp(\ii\omega t)
\end{align}
and define the \emph{total Lagrangian} (in the frequency domain) $\mathcal{L}_A$:
\begin{align}
 &\mathcal{L}_A=\tilde{\mathcal{L}}+\mathcal{L}, &\begin{cases}
\mathcal{\tilde{L}}=\tilde{\mathbf{D}}_0\tilde{\mathbf{E}}^*-\tilde{\mathbf{H}}\tilde{\mathbf{B}}^*\\
\mathcal{L}=(\mathbf{D}_0+\mathbf{P}^{\mathrm{NL}}/2)\mathbf{E}^*-\mathbf{H}\mathbf{B}^*
\end{cases} 
\end{align}
where $\mathcal{\tilde{L}}$ and $\mathcal{L}$ are the Lagrangian fields modeling  the linear canonical structure and the whole nonlinear medium, respectively; the former contains:
\begin{align}
 &\tilde{\mathbf{D}}_0=\vv{\tilde{\epsilon}}\tilde{\mathbf{E}}, &\tilde{\mathbf{B}}=\vv\mu\tilde{\mathbf{H}},
\nonumber\\
&\tilde{\mathbf{E}}=-\nabla\tilde{\Phi}-\ii\omega\tilde{\mathbf{A}}, &\tilde{\mathbf{B}}=\nabla\times\tilde{\mathbf{A}}
\end{align}
and the latter:
\begin{align}
&\mathbf{D}_0=\vv\epsilon\mathbf{E}, & \mathbf{B}=\vv\mu\mathbf{H},
\nonumber\\
&\mathbf{E}=-\nabla\Phi-\ii\omega\mathbf{A}, &\mathbf{B}=\nabla\times\mathbf{A},
\nonumber\\
&\mathbf{P}^{\mathrm{NL}}=\vv\de\mathbf{E}
\end{align}
with $\de_{ij}=\epsilon_0\chi^{(3)}_{1111}(\mathbf{r};\omega)E_i^*E_j(\delta_{ij}+1/2)/2$ \cite{NO}. From the group of rotations $\mathbf{v}=\tilde{A_j}\partial/\partial A^j-A_j\partial/\partial \tilde{A}^j+\tilde{\Phi}\partial/\partial \Phi-\Phi\partial/\partial \tilde{\Phi}+\mathrm{c.c.}$, through Noether's symmetries, we obtain the conservation law:
\begin{equation}
  \label{orto0}
  \nabla(\tilde{\mathbf{E}}^*\times\mathbf{H}+\mathbf{E}\times\tilde{\mathbf{H}}^*)+\ii\omega(\epsilon_0\vv{\Delta\epsilon}+\mathbf{P}^{\mathrm{NL}})\tilde{\mathbf{E}}^*=0
\end{equation}
which can be regarded as a generalization of the Lorentz reciprocity theorem. If we apply the same procedure to the canonical structure only (i.e., for $\vv{\Delta\epsilon}=\mathbf{P}^{\mathrm{NL}}=0$), we obtain:
\begin{equation}
  \label{orto1}
  \nabla(\tilde{\mathbf{E}}^*\times\tilde{\mathbf{H}}'+\tilde{\mathbf{E}}'\times\tilde{\mathbf{H}}^*)=0
\end{equation}
where $(\tilde{\mathbf{E}},\tilde{\mathbf{H}})$ and $(\tilde{\mathbf{E}}',\tilde{\mathbf{H}}')$ are two sets of solutions. The conservation laws (\ref{orto0})-(\ref{orto1}) support the expansion of the fields $(\mathbf{E},\mathbf{H})$ into an orthonormal eigenbasis provided by the canonical structure, resulting in a series of first-order differential equations completely equivalent to Maxwell's equations. In particular Eq. (\ref{orto1}) yields the orthogonality between eigenvectors while Eq. (\ref{orto0}) allows us to calculate the expansion coefficients via orthogonality. The canonical eigensolutions are of the form:
\begin{align}
\begin{bmatrix}
\mathbf{E}_\nu\\
\mathbf{H}_\nu
\end{bmatrix}=\bigg( \begin{bmatrix}
\mathbf{E}_{\nu t}\\
\mathbf{H}_{\nu t}
\end{bmatrix}
+\mathbf{\hat{z}}
\begin{bmatrix}
E_{\nu z}\\
H_{\nu z}
\end{bmatrix}
\bigg)\exp(\ii\beta_\nu z)  
\end{align}
with subscript $t$ denoting the transverse component, $\beta_\nu$ being the propagation eigenvalue, the spectrum of which encompasses a discrete real set (bound states or guided modes), a continuous real set (unbound states or radiation modes) and a continuous purely imaginary set (evanescent states) \cite{IO}. The orthogonality between the eigenstates is expressed by the density conservation of (\ref{orto1}) along $z$:
\begin{align}
  \label{orto2}
\mathcal{I}_{\nu\mu}=\iint dxdy (\tilde{\mathbf{E}}_{\nu t}^*\times\tilde{\mathbf{H}}_{\mu t}'&+\tilde{\mathbf{E}}_{\mu t}'\times\tilde{\mathbf{H}}^*_{\nu t})=\frac{-\beta_\nu\delta(\nu-\mu)}{\lvert \beta_\nu\rvert}
\end{align}
Through (\ref{orto2}), the vector field $(\mathbf{E},\mathbf{H})$ can be expanded by employing the transverse portion of the canonical eigenvectors, the longitudinal components following from Maxwell's equations. To perform this expansion, we begin by writing the transverse portions of the electromagnetic fields as:
\begin{align}
  \label{expp0}
&\begin{bmatrix}
\mathbf{E}_t\\
\mathbf{H}_t
\end{bmatrix}=\int d\nu C_\nu(z)\begin{bmatrix}
\tilde{\mathbf{E}}_{\nu t}\\
\tilde{\mathbf{H}}_{\nu t}
\end{bmatrix}
e^{\ii\beta_\nu z},
\end{align}
with envelopes $C_\nu$ varying in the direction of propagation $z$. We then write the Euler-Lagrange equation of motions corresponding to the Lagrangian coordinate $A_z$:
\begin{align}
\label{edc0}
\nabla_t\times \mathbf{H}_t=\ii\omega\epsilon_0\big[(\epsilon^{'}_{zz}+\de_{zz})E_z+\de_{zt}\cdot\mathbf{E}_t\big]=0
\end{align}
being $\de_{zt}$ a vector of components $\de_{zt}=[\de_{zx},\de_{zy}]$ and having expanded $1/(\epsilon'+\de_{zz})=(1-\de_{zz}/\epsilon')/\epsilon'+O(\de_{zz}^2)$ to first order in $\de_{zz}$ (since $\de_{zz}\ll \epsilon'$ \cite{bf_npx_ciatta}). We find the expression of $E_z$ by direcly solving (\ref{edc0}), keeping into account the transverse modal expansion (\ref{expp0}):
\begin{align}
  E_z=\int d\nu C_\nu(z)\bigg[\frac{\tilde{E}_{\nu z}\te}{\epsilon'}\bigg(1-\frac{\de_{zz}}{\epsilon'}\bigg)+&\frac{\de_{zt}\tilde{\mathbf{E}}_{\nu t}}{\epsilon'}\bigg]
\nonumber\\
&\times e^{\ii\beta_\nu z},
\end{align}
being $\tilde{E}_{\nu z}=\frac{\nabla_t\times\vv{\tilde{H}}_{\nu t}}{\ii\omega\epsilon_0\te}$. We then write the $z-$component of Eq. (\ref{edcm0}):
\begin{align}
  \nabla_t\times\mathbf{E}_t=-\ii\omega\mu H_z,
\end{align}
and solve for $H_z$, keeping into account Eqs. (\ref{expp0}), thus obtaining:
\begin{align}
  H_z=\int d\nu C_\nu(z)\tilde{H}_{\nu z}
\end{align}
being $\tilde{H}_{\nu z}=\ii\frac{\nabla_t\times\vv{\tilde{E}}_{\nu t}}{\omega\mu}$. Summing up, our expansion in transverse modes is:
\begin{align}
  \label{expp01}
&\begin{bmatrix}
\mathbf{E}_t\\
\mathbf{H}_t
\end{bmatrix}=\int d\nu C_\nu(z)\begin{bmatrix}
\tilde{\mathbf{E}}_{\nu t}\\
\tilde{\mathbf{H}}_{\nu t}
\end{bmatrix}
e^{\ii\beta_\nu z},
\end{align}
\begin{align}
\begin{bmatrix}
\label{expp12}
E_z\\
H_z
\end{bmatrix}=\int d\nu C_\nu(z)&\begin{bmatrix}
\frac{\tilde{E}_{\nu z}\te}{\epsilon'}(1-\frac{\de_{zz}}{\epsilon'})+\frac{\de_{zt}\tilde{\mathbf{E}}_{\nu t}}{\epsilon'}\\
\tilde{H}_{\nu z}
\end{bmatrix}e^{\ii\beta_\nu z}.
\end{align}
The integrals (\ref{expp0})-(\ref{expp12}) should be calculated over the whole spectrum of $\beta$; however, since the power $P$ carried by each eigenstate is $P=\mathcal{I}_{\nu\nu}/4$ \cite{IO}, we are interested here in the evolution of the bound states (i.e. modes with $P\in\Re$ and well confined within the canonical structure). The latter evolution is obtained as follows. We first integrate Eq. (\ref{orto0}) across the plane $(x,y)$:
\begin{align}
\label{intxu}
  \iint dxdy \bigg[\frac{\partial}{\partial z}&\big(\tilde{\mathbf{E}}^*\times\mathbf{H}+\mathbf{E}\times\tilde{\mathbf{H}}^*\big)\cdot\hat{\mathbf{z}}\bigg]+\nonumber\\
&\iint dxdy \bigg[\ii\omega(\epsilon_0\vv{\Delta\epsilon}+\mathbf{P}^{\mathrm{NL}})\tilde{\mathbf{E}}^* \bigg]=0.
\end{align}
We then substitute Eqs. (\ref{expp01})-(\ref{expp12}) into (\ref{intxu}). The contribution of the first integral of (\ref{intxu}), keeping into account the normalization condition (\ref{orto2}), is:
\begin{align}
  \int dxdy&\big(\tilde{\mathbf{E}}_{tn}^*\times\tilde{\mathbf{H}}_{tm}+\tilde{\mathbf{E}}_{tn}\times\tilde{\mathbf{H}}_{tm}^*\big)\nonumber\\
&\times\bigg[\frac{\partial}{\partial z}C_m+\ii(\beta_m-\beta_n)\bigg]=-\frac{\partial}{\partial z}C_n,
\end{align}
where we employed the indices $n$ and $m$ to indicate the canonical eigenvectors of the system. The substitution of Eqs. (\ref{expp01})-(\ref{expp12}) into the second integral of (\ref{intxu}), after some straighfroward algebra, yields to the DNLS equation:
\begin{align}
  \label{ssdnls0}
\ii\frac{\partial}{\partial \xi}\psi_n+\psi_{n+1}+\psi_{n-1}+\lvert\psi_n\rvert^2\psi_n=0,
\end{align}
where we introduced the dimensionless $\xi=z K_{n,n+1}$ and:
\begin{align}
\psi_n=C_n\sqrt{\frac{\omega\epsilon_0\chi^{(3)}_{1111}\iint dxdy \mathbf{P}^{\mathrm{N}}\tilde{\mathbf{E}}_n}{2K_{n,n+1}}}\exp\bigg(-\ii \frac{K_{n,n}\xi}{K_{n,n+1}}\bigg)  
\end{align}
with:
\begin{align}
K_{n,n+1}&=\omega\epsilon_0\iint dxdy
\nonumber\\
&\times\bigg(\vv{\Delta\epsilon}\tilde{\mathbf{E}}_{n,t}\tilde{\mathbf{E}}_{n+1,t}+\frac{\Delta\epsilon\tilde{\epsilon}\tilde{E}_{n,z}\tilde{E}_{n+1,z}}{\epsilon'}\bigg)
\nonumber\\
\mathbf{P}^{\mathrm{N}}&=\vv{\tilde{\de}}\tilde{\mathbf{E}}_n+\mathbf{\hat{z}}\Delta\epsilon
\nonumber\\
&\times\frac{\mathbf{\tilde{\de}}_{zx}\tilde{\mathbf{E}}_{n,x}+\mathbf{\tilde{\de}}_{zy}\tilde{\mathbf{E}}_{n,y}-\te\tilde{\de}_{zz}\tilde{E}_{n,z}/\epsilon'}{\epsilon'}
\end{align}
where $\tilde{\de}_{ij}=\sigma_i\sigma_j(\delta_{ij}+1/2)$, $[\sigma_x,\sigma_y,\sigma_z]=[\tilde{E}_{n,x},\tilde{E}_{n,y},\te\tilde{E}_{n,z}/\epsilon']$ and  $n+h$ denotes the bound state of the canonical structure centered in $y=h\Lambda$. It is worth stressing that the reduction procedure based upon symmetries of Maxwell's electrodynamics is exact; the only assumption used in deriving Eq. (\ref{ssdnls0}), in spite of the OR, is that guided modes are well confined. \emph{The latter hypothesis, stating that $\beta_n\neq\beta_m$ $\forall m,n$ \cite{bf_nkm_crosia}, allows us to neglect mismatched terms given by bound/unbound/evanescent states proportional to $\exp[\ii(\beta_{n}-\beta_{m})z]$ and guided modes with $n+h\geq n+2$}. The latter turns possible due to the exponential dependence of $K_{mn}$ on the the distance between sites $n$ and $m$ \cite[see e.g.][Chapt. 13, page 522, Eq. 13.8-9]{YAR}, yielding an appreciable contribution only for nearest-neighbor channels $n$ and $n\pm 1$.\\ 
 This analysis demonstrates that the discrete nonlinear Schr\"odinger equation models the dynamics in the general case of discrete nonparaxial vector propagation in lattices with a nonlinear tensorial response.
\section{Classification analysis}
The properties of the DNLS are established by its structure. In particular\\
i) The two invariants of motion:
\begin{align}
&W=\sum_n \lvert\psi_n\rvert^2,\nonumber\\
&H=\sum_n \psi_n\psi_{n+1}^*+\psi_n^*\psi_{n+1}+\frac{1}{2}\lvert\psi_n\rvert^4  
\end{align}
where $W$ physically corresponds to the power and $H$ to the Hamiltonian. These quantities, generated by the  symmetries $\{\mathbf{v}_1=\sum_n \ii\psi_n\partial/\partial \psi_n+\mathrm{c.c.},\mathbf{v}_2=\partial/\partial z\}$ of the Lagrangian $\mathcal{L}=\sum_n\frac{1}{2}(\ii\psi_n^*\frac{\partial\psi_n}{\partial z}+\mathrm{c.c.})-H$, characterize nonlinear waves \cite{rev_wa_TrillA} as well as chaos \cite{ds_chs_herbsa}.\\
ii) The existence of linear plane-wave solutions $\psi_n=\exp(\ii\beta z-\ii q n)$ with a (real) periodic dispersion relation $\beta=2\cos q$ \cite{SS}.\\
To obtain the simplest differential equation with the symmetries expressed by $\{\mathbf{v}_1,\mathbf{v}_2\}$ and compatible with ii) firstly we have to solve a classification problem, which yields all the equations with given symmetries. To this  extent we generalize the theory developed in \cite{dis_cp_doroda} for difference schemes. Specifically, we start by considering a general system of differential-difference equations $\mathbf{Q}=[Q^1,Q^2,Q^3]=0$ involving nearest-neighbor (discrete) interactions:
\begin{align}
\label{guess}
  \begin{bmatrix}
Q^1\\
Q^2\\
Q^3
\end{bmatrix}=
\begin{bmatrix}
\frac{\partial \psi}{\partial \eta}\\
\frac{\partial \phi}{\partial \eta}\\
d_+
\end{bmatrix}-\begin{bmatrix}
f^{1}(\eta,x,d_-,[\psi,\phi],[\psi_+,\phi_+],[\psi_-,\phi_-])\\
f^{2}(\eta,x,d_-,[\psi,\phi],[\psi_+,\phi_+],[\psi_-,\phi_-])\\
f^{3}(\eta,x,d_-,[\psi,\phi],[\psi_+,\phi_+],[\psi_-,\phi_-])
\end{bmatrix}
\end{align}
with unknown functions $f^i$ ($i\in [1,3]$) depending up on the discrete variables $[\psi(x_n,\eta),\phi(x_n,\eta)]=[\psi,\phi]\equiv[\psi,\psi^*]$ and their neighbors $[\psi(x_{n\pm 1},\eta),\phi_(x_{n\pm 1},\eta)]=[\psi_{\pm},\phi_{\pm}]\equiv[\psi_\pm,\psi_\pm^*]$, defined in discrete $[x,d_+,d_-]=[x_n,x_{n+1}-x_n,x_n-x_{n-1}]$ and continuous $\eta=\ii\xi$ spaces of independent variables. The first two of (\ref{guess}) describe the evolution of $\psi$ and $\phi$ and the third defines the lattice. The introduction of the complex space $\eta$ guarantees that $\psi and \phi$  be real and the classification problem to be solved in the real domain $\Re$. To perform the classification, we need to find the structure of $f^i$ supporting the symmetries generated by the basis $\{\mathbf{v}_1=\partial/\partial \eta,\mathbf{v}_2=\psi\partial/\partial\psi-\phi\partial/\partial\phi \}$ and satisfying a real dispersion relation. Each of the symmetries possessed by (\ref{guess}) should satisfy the invariance criterion:
\begin{align}
\label{sinvo0}
 &\mathbf{pr}^{1}\mathbf{v}_i Q^{j}\lvert_{Q^{j}=0}=0, &\begin{cases}
i\in[1,2]\\
j\in[1,2,3]
\end{cases}
\end{align}
with $\mathbf{pr}^{1}$ the prolongation operator (of order 1) defined as:
\begin{align}
\label{prcp0}
 &\mathbf{pr}^{1}\mathbf{v}=\mathbf{v}+ \zeta(x_+,y_+)\frac{\partial}{\partial x_+}+\zeta(x_-,y_-)\frac{\partial}{\partial x_-}+\sigma_i(x_+,y_+)
\nonumber\\
&\times \frac{\partial}{\partial \gamma^{i}_+}+\sigma_i(x_-,y_-)\frac{\partial}{\partial \gamma^{i}_-}+\omega_i\frac{\partial}{\partial \gamma_\eta^i}
\end{align}
for a generic vector field:
\begin{align}
\mathbf{v}=\zeta\frac{\partial}{\partial x}+\tau\frac{\partial}{\partial\eta}+\sigma_i\frac{\partial}{\partial\gamma^i}  
\end{align}
with:
\begin{align}
&\gamma=[\psi,\phi],
\nonumber\\
&\gamma_{\pm}=[\psi_\pm,\phi_\pm],
\nonumber\\
& \omega_i=\bigg[\frac{\partial\sigma_i}{\partial\eta}-\frac{\partial\zeta}{\partial\eta}\frac{\partial\gamma_i}{\partial x}-\frac{\partial\tau}{\partial\eta}\frac{\partial\gamma_i}{\partial \eta}\bigg],
\nonumber\\
 &\gamma_\eta^i=\frac{\partial \gamma^i}{\partial\eta} 
\end{align}
The substitution of (\ref{guess}) and (\ref{prcp0}) into (\ref{sinvo0}) results in a system of equations which can be solved on the characteristics to yield the functional form of $f^i$ ($i\in [1,3]$):
\begin{align}
\label{guess1}
 \begin{bmatrix}
\frac{\partial \psi}{\partial \eta}\\
\frac{\partial \phi}{\partial \eta}\\
d_+
\end{bmatrix}=\begin{bmatrix}
\psi f^1(x,d_-,\psi\phi,\psi\phi_\pm,\phi\psi_\pm,\frac{\phi_\pm}{\phi},\frac{\psi_\pm}{\psi})\\
-\phi f^{2}(x,d_-,\psi\phi,\psi\phi_\pm,\phi\psi_\pm,\frac{\phi_\pm}{\phi},\frac{\psi_\pm}{\psi})\\
d_-f^{3}(x,d_-,\psi\phi,\psi\phi_\pm,\phi\psi_\pm,\frac{\phi_\pm}{\phi},\frac{\psi_\pm}{\psi})
\end{bmatrix}
\end{align}
The simplest differential-difference equation encompassed by (\ref{guess1}) is obtained for $f^{3}=1$ (i.e., for $d_+=d_-=d$) and for $f^1,f^2$ depending on just one nonlinear product $\psi\phi$ and one couple of linear terms $[\psi_\pm/\psi,\phi_\pm/\psi]$ , the latter giving a real dispersion relation as required by ii):
\begin{align}
\label{send0}
&\frac{\partial \psi}{\partial \eta}=\frac{\psi_{+}+\psi_{-}+\psi^2\phi}{d},
&\frac{\partial \phi}{\partial \eta}=-\frac{\phi_{+}+\phi_{-}+\phi^2\psi}{d}
\end{align}
The factor $1/d$ guarantees the existence of the continuous limit of (\ref{send0}) for $d\rightarrow 0$, $[\psi(\eta),\phi(\eta)]=d[u(x,\eta),v(x,\eta)]$ and $x=nd$. However, in our dimensionless example we can always set $d=1$. Finally, going back to the original variables $[\xi,\psi,\psi^*]$, we obtain the DNLS equation (\ref{ssdnls0}). The solution of the classification problem therefore demonstrates that the discrete nonlinear Schr\"odinger equation has the simplest structure compatible with its properties.\\ 
In summary our top-down/bottom-up analysis shows that:\\
I) The DNLS is ``universal'' because it is able to model discrete nonparaxial vector propagation in lattices with a tensorial nonlinearity. This stems from fundamental principles and exploits the conservation laws arising from Noether's symmetries [Eqs. (\ref{orto0})-(\ref{orto1})].\\
II) The DNLS is ``simple'' because it contains the smallest number of terms that provide a functional form compatible with its properties. This stems from solving the classification problem [Eqs. \ref{guess1}] by means of a Lie-symmetry analysis.\\
We can therefore state that \emph{the DNLS equation is a general model, the importance of which goes far beyond the limits within which it is usually derived, being the simplest model able to describe nonparaxial vector energy propagation in a discrete medium with a tensorial nonlinear response}. It is worth underlining that this result was obtained by simply exploiting symmetries, i.e. a general approach that can be extended to all the physical contexts, where yet model ``universality'' has not been assessed.\emph{ To provide an example we consider the case of two (or more) dimensional optical lattices (see e.g. \cite{vs_ol_bartaa,gs_td_efrema,rev_wa_FleisAA}). Our theory, although applied in the present context to the case of one dimensional optical lattices, can be straightforwardly generalized to two (or even more) dimensional systems, allowing to investigate the universality of higher dimensional physical models.}
\\
\section{Conclusions}
In conclusion we undertook a general procedure to establish both the significance and the ``universality'' of equations describing physical events, attributing such features to those which model the most general dynamics in the most elementary form possible. We demonstrated our approach in the case of optical lattices, revealing that the discrete nonlinear Schr\"odinger equation, previously considered in the specific context of scalar, paraxial propagation in weakly modulated media, is far more ``universal'' and important as it describes discrete nonparaxial vector energy propagation in materials with a tensorial Kerr response. The general character of the outlined procedure as well as the contemporary interest of discrete nonlinear systems are expected to contribute towards a better comprehension of model validity and applicability in physics.
\section{Acknowledgments}
The authors are grateful to F. T. Arecchi and T. Minzoni for stimulating discussions.


\end{document}